# Manual do Software EnergySaver

**Davi Guimarães da Silva, Marla Teresinha Barbosa Geller, Dalton Felipe Silva Varão, João Bentes, Mauro Sérgio dos Santos Moura, Yasmin Braga Teixeira, Clayton André Maia dos Santos, Anderson Alvarenga de Moura Meneses.**

*Programa de Pós-graduação Doutorado em Sociedade, Natureza e Desenvolvimento,*
*Laboratório de Inteligência Computacional - LABIC*
*Universidade Federal do Oeste do Pará – UFOPA - Santarém – PA – Brasil*

Autor do Manual:
Davi Guimarães da Silva
Programa de Pós-graduação Doutorado em Sociedade, Natureza e Desenvolvimento, Laboratório de Inteligência Computacional - LABIC, Universidade Federal do Oeste do Pará – UFOPA - Santarém – PA – Brasil.
Email: davigsmsc@gmail.com

**Resumo:** Eficiência energética é um tópico que têm atraído a atenção de pesquisadores nos últimos anos, no sentido de buscar soluções de sustentabilidade para a produção de energia e redução de seus custos, visando proporcionar o equilíbrio entre desenvolvimento e proteção dos recursos naturais. Com base nisso, propusemos o software EnergySaver que tem por objetivo o monitoramento do consumo de energia elétrica, desde a captação de dados até a previsão de consumo para o mês seguinte. Para criação do Energy Saver, utilizou-se tecnologias Open Source aplicadas a internet das coisas (IoT), sistemas embarcados e Redes Neurais Long Short-Term Memory (LSTM). Porém, para haver sintonia entre os pesquisadores atuais e os que futuramente possam manipular este software, torna-se essencial a criação de um Manual do Software, onde todos os detalhes da sua implementação são descritos minuciosamente. Assim sendo, este artigo descreve todos os passos para a implantação do sistema, desde o esquema metodológico do sistema, a sua modelagem com UML, até os módulos que o compõem, tornando-se um Manual de utilização do mesmo.

## Introduction

A energia elétrica possui uma grande importância para a vida dos indivíduos, empresas, instituições, e está diretamente ligada ao desenvolvimento dos países. Na atualidade, é indiscutível a dependência extrema desse recurso pois a realização de diversas atividades econômicas e o bem estar social estão diretamente ligados ao consumo de energia elétrica.

Para com Goldemberg (2003), o consumo de energia é um índice representativo do acesso da população às condições básicas de vida. Porém, é importante destacar que a conquista da qualidade de vida não deve comprometer a integridade do planeta, ou seja, são necessárias atitudes sustentáveis de modo que seja possível manter a comodidade social adquirida e também minimizar os danos aos recursos naturais, valendo-se de ações que promovam a eficiência energética.

Dentre várias pesquisas que buscam viabilizar o uso eficiente de energia elétrica, destaca-se o uso de sistemas de Internet das Coisas (Internet of Things – IoT). De acordo com Borgia (2014), a IoT trata da capacidade dos dispositivos de se comunicarem entre si, mediante uso de sensores e atuadores, a fim de monitorar o ambiente. Outro conceito importante a se destacar e que pode ser aplicado ao uso eficiente de energia elétrica é a Aprendizagem profunda (Deep learning). Trata-se de um conjunto de técnicas que utilizam redes neurais artificiais profundas, com muitas camadas intermediárias entre a camada de entrada e a de saída. De modo geral, ela treina computadores para realizar tarefas como seres humanos. (LECUN et al., 2015). Vale ressaltar também a importância de fazer a modelagem de qualquer sistema, evitando assim grandes erros durante sua implementação, e para isso utilizou-se a linguagem de modelagem unificada (UML – Unified Modeling Language).

Assim sendo, Baumgartner (2021) destaca que faz-se necessário contar com documentos que descrevem o funcionamento, configurações e forma de operação das aplicações para o ambiente que foram projetados para que os interessados possam sanar dúvidas de forma adequada. Nesse sentido, é essencial a criação de um manual, pois trata-se um documento que tem por objetivo principal ajudar os utilizadores a entender como um produto funciona e como utilizá-lo, além de oferecer auxílio para identificar e resolver problemas, traduz conceitos técnicos para uma linguagem simples que todos os interessados possam compreender, sendo essencial no processo de desenvolvimento e utilização de software.

O artigo está organizado da seguinte forma: Na Seção II será apresentado o sistema de modo geral. A seção III detalha o processo de modelagem do EnergySaver. A Seção IV detalha o móduko de previsão para o sistema . Na seção V são tecidas as considerações gerais e, por fim, as Referências utilizadas.

## Conhecendo o Sistema

O EnergySaver é um software que tem por objetivo o monitoramento de energia elétrica de aparelho de ar-condicionado da Universidade Federal do Oeste do Pará. O sistema de captação de dados utiliza tecnologias Open Source, Conceitos de Internet das Coisas e sistemas embarcados.

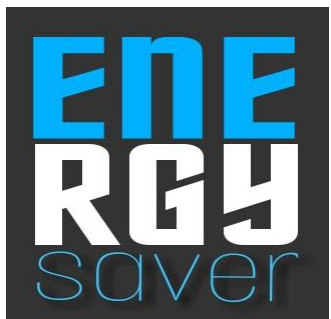

**Fig. 1:** Logo do Sistema

Este sistema pode ser utilizado em ambientes Windows e Linux, que a partir de configurações básicas (descritas nas próximas seções) pode ser instalado e executado tanto no Raspberry Pi3 quanto no Servidor Local.

*Sobre este manual*

Este manual bem como o sistema em questão, foram desenvolvidos completamente pela equipe do Projeto Energy Saver na Universidade Federal do Oeste do Pará (UFOPA), em pesquisa desenvolvida no Laboratório de Inteligência Computacional (LABIC), coordenada e orientada pelo Professor Dsc. Anderson Alvarenga de Moura Meneses.

*Sobre o uso e distribuição do software*

O Energy Saver é um software livre , criado totalmente com ferramentas OpenSource e sua utilização pode ser autorizada apenas pelo coordenador do LABIC para fins de pesquisa.

*Esquema metodológico do sistema*

De modo geral, o esquema metodológico do sistema é da seguinte forma:

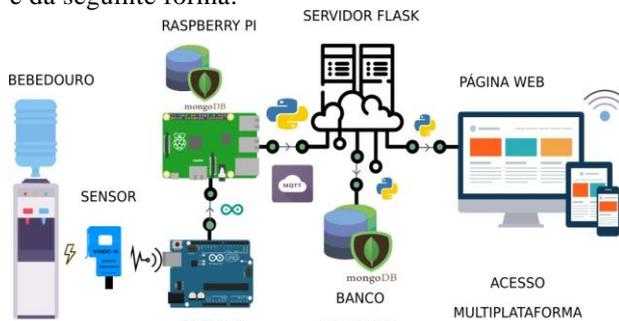

**Fig. 2:** Esquema metodológico do sistema

Inicialmente os dados de consumo são capturados pelo sensor e arduino, sendo enviados em seguida para o Raspberry Pi3, que por sua vez envia pelo mosquitto broker (MQTT) ao servidor Flask. No servidor, esses dados são armazenados no banco de dados MongoDB e também exibidos em tempo real em uma página web existente no Servidor Local. Essa página pode ser acessada via endereço de ip do servidor e porta 5000 de qualquer máquina que esteja conectada à rede local/wifi.

*Função de cada parte do sistema*

Cada equipamento utilizado no EnergySaver será descrito de modo geral a seguir:

a) **Sensor de corrente elétrica com arduino:** O medidor de energia consiste basicamente em um sensor de corrente e um de tensão conectados a um circuito microprocessado, que no caso é o Arduino, conforme figura 3.

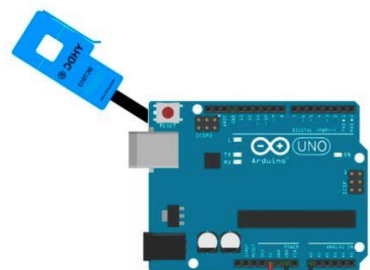

**Fig. 3:** Sensor com Arduino

Fazemos a leitura das duas variáveis elétricas básicas e a partir delas determinamos a potência elétrica instantânea, bem como a energia consumida em um determinado intervalo de tempo.

b) **Raspberry Pi3:** O Raspberry Pi é um miniPC que tem o tamanho de um cartão, conforme figura 4.

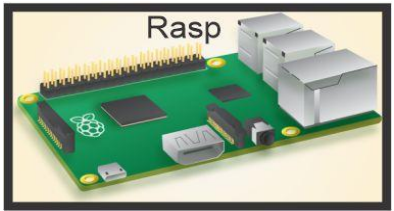

**Fig. 4:** Raspberry Pi

A sua placa é um computador barato, portátil e versátil, usado principalmente em projetos de programação, robótica e em iniciativas em geral com software e hardware livre. Ele é usado, também, para automação doméstica e outras aplicações envolvendo Internet das Coisas.

c) *Mosquitto:* O Mosquitto é um intermediário de mensagens de código-fonte aberto (licenciado por EPL / EDL) que implementa as versões 3.1 e 3.1.1 do protocolo MQTT. Esse protocolo fornece um método leve de executar mensagens usando um modelo de publicação / assinatura. Isso o torna adequado para mensagens da Internet das Coisas, como sensores de baixa potência ou dispositivos móveis, bem como telefones, computadores embutidos ou microcontroladores.

d) *Server NodeJS:* Escrito em Python e disponível sob licença BSD (Licença de código aberto), o Flask é um micro-framework multiplataforma que provê um modelo simples para o desenvolvimento web, sendo bastante utilizado para criação de microsserviços, como APIs RESTful. O Flask depende de duas bibliotecas externas, Werkzeug e Jinja2. Werkzeug é um toolkit para WSGI, a interface padrão entre aplicações web Python e servidores HTTP para desenvolvimento e implantação. Jinja2 renderiza templates.

e) *MongoDB:* O MongoDB é um banco de dados orientado a documentos (document database) no formato JSON, ou seja, diferente de um banco de dados relacional, ele não possui como restrição a necessidade de ter as tabelas e colunas criadas previamente, permitindo que um documento represente toda a informação necessária, com todos os dados que você queira.

f) *Micro-Framework Flask:* O Flask oferece sugestões, mas não impõe nenhuma dependência ou layout do projeto. Cabe ao desenvolvedor escolher as ferramentas e bibliotecas que deseja usar. Existem muitas extensões fornecidas pela comunidade que facilitam a adição de novas funcionalidades.

g) *Aplicação:* A aplicação utiliza como base o micro-framework flask sendo que sua atualização de dados ocorre em tempo real via protocolo MQTT pelo Raspberry Pi. Os dados recebidos podem ser baixados a qualquer momento em formato CSV diretamente na aplicação.

## Processo de Modelagem do Energy Saver utilizando UML

O processo de modelagem do sistema Energy Saver foi realizado a partir de Ferramentas e processos, que serão descritos nas seções seguintes.

*Ferramentas e processo para modelagem do EnergySaver*

A linguagem de modelagem unificada (UML – Unified Modeling Language) é a notação gráfica comumente adotada na maioria dos processos de desenvolvimento.

UML é uma linguagem genérica que serve para modelagem de diversos tipos de software. Ela possui diversos tipos de diagramas, apoiando a criação de diferentes modelos para diferentes perspectivas do sistema. Mais informações sobre UML podem ser obtidas no guia de uso desenvolvido pelos próprios criadores desta linguagem, sendo bem completo para que o leitor entenda como fazer sua utilização, bem como a aplicação prática.

Neste manual de software, para o desenvolvimento de alguns diagramas UML, foi utilizada a ferramenta Astah, tendo em vista ser bastante difundida no meio acadêmico, sendo também de fácil manipulação. A modelagem do sistema foi feita em Geller (2021).

O Processo para modelagem do sistema Energy Saver foi adaptado de Reggio (2018) que utiliza a UML para especificação dos requisitos funcionais e não funcionais de sistemas IoT, seguindo o paradigma orientado a serviços (SOA).

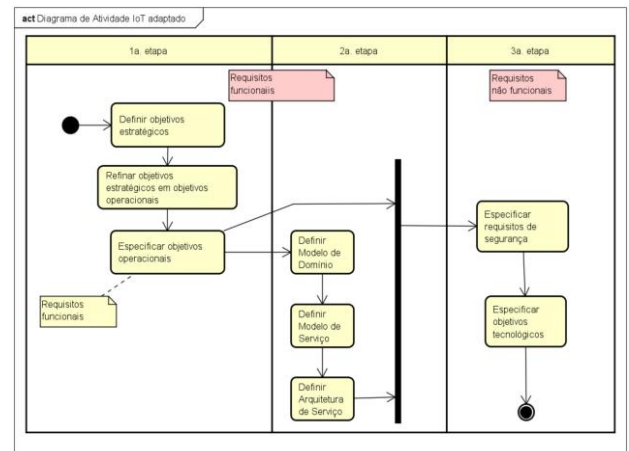

**Fig. 5:** Customização do processo para modelagem IoT.

A figura 5 apresenta as três etapas do método que produzem como artefatos o modelo de domínio, modelo de serviços, arquitetura dos serviços e o modelo de segurança. Essas etapas serão descritas a seguir.

*Primeira Etapa*

Nessa etapa é realizada a definição dos objetivos estratégicos, bem como seu refinamento em objetivos operacionais, da seguinte forma:

a) **Definição dos objetivos estratégicos:** Na primeira etapa a definição do objetivo e dos subobjetivos estratégicos é representada pelos diagramas de casos de uso, como mostra a figura 6.

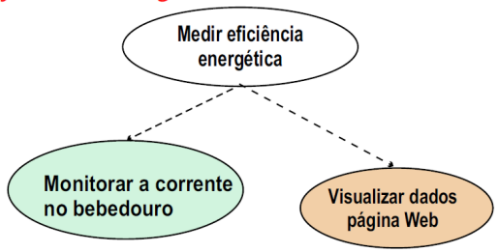

**Fig. 6:** Objetivos estratégicos e sub objetivos estratégicos.

O objetivo geral do sistema é "medir a eficiência energética", que dá origem aos subojetivos estratégicos, sendo eles "monitorar a corrente elétrica" e "visualizar dados na página Web".

b) **Refinamento dos objetivos estratégicos em objetivos operacionais:** Os subobjetivos estratégicos são refinados em outros casos de uso dando origem aos objetivos operacionais que se constituem nas funcionalidades do sistema, como mostram as figuras 7 e 8.

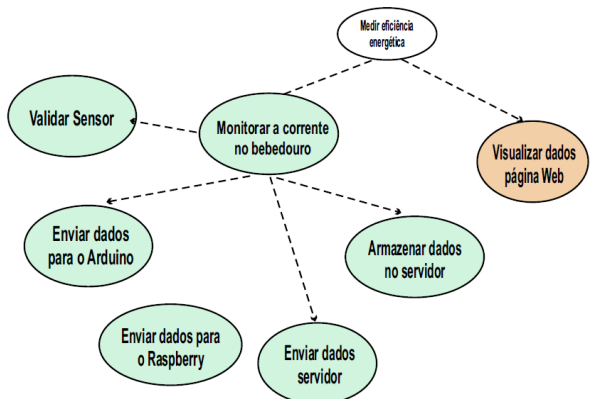

**Fig. 7:** Objetivos operacionais relativos ao caso de uso "Monitorar a corrente do bebedouro".

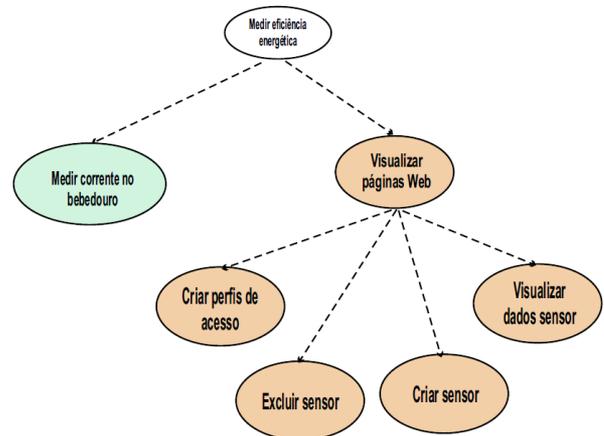

**Fig. 8:** Objetivos operacionais relativos ao caso de uso "Visualizar páginas Web"..

*Segunda Etapa*

Nessa etapa são apresentados os modelos de domínio, arquitetura dos serviços, modelo de segurança e identificação dos objetivos tecnológicos, da seguinte forma:

a) **Modelo de Domínio:** O modelo de domínio mostra a visão estática do sistema, que é representada pelo diagrama de classes. Essas classes são estereotipadas como <<participant>>, como mostra a figura 9.

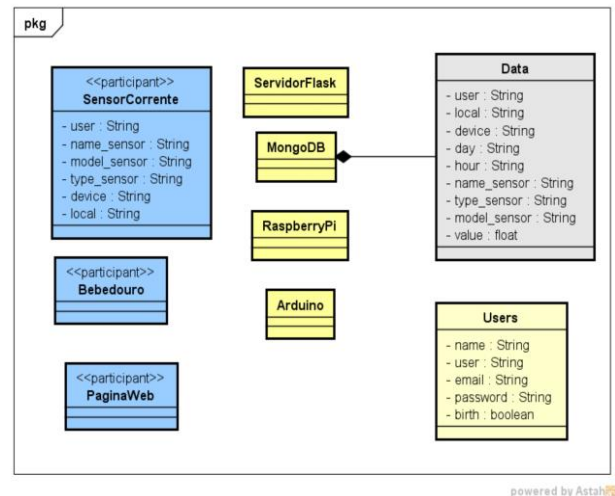

**Fig. 9:** Objetivos estratégicos e sub objetivos estratégicos.

Ainda na segunda etapa é criado o modelo de serviço a exemplo do serviço de atualização, que é representado pela classe Atualiza com suas interfaces AtualizaIn e AtualizaOut, representada na figura 10.

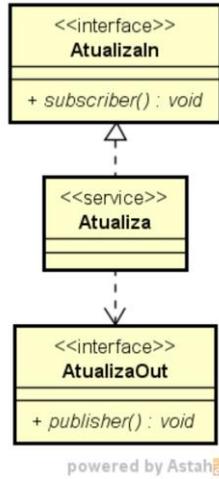

**Fig. 10:** Classes que participam do serviço Atualiza.

A troca de mensagens para realização desse serviço é mostrada no diagrama de sequência da figura 11 e os dados atualizados são representados pela classe Data na figura 12.

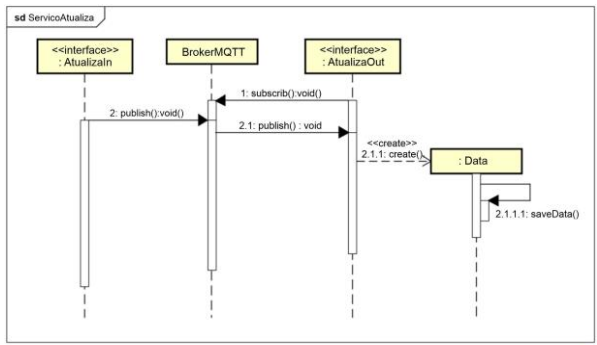

**Fig. 11:** Diagrama de sequência do serviço Atualiza.

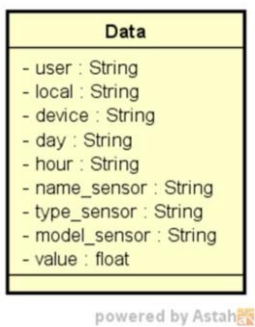

**Fig. 12:** Classe Data.

b) ***Arquitetura dos serviços:*** A arquitetura do serviço utiliza o diagrama de componentes, como mostra a figura 13, com suas interfaces de comunicação publischer/subscriber representadas.

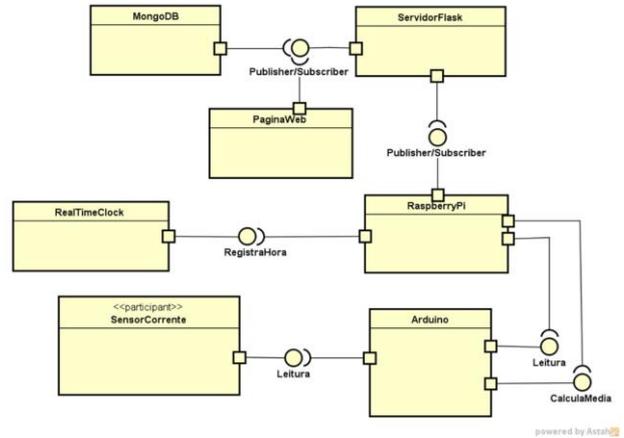

**Fig. 13:** Diagrama de componentes representando a arquitetura dos serviços.

*Terceira Etapa*

Nessa etapa são apresentados os modelos de segurança e identificação dos objetivos tecnológicos, da seguinte forma:

a) ***Modelo de segurança:*** A figura 14 representa os componentes do sistema distribuídos em suas diferentes camadas, ressaltando a necessidade de dispositivos de segurança entre uma camada e outra, como especificado na figura 15.

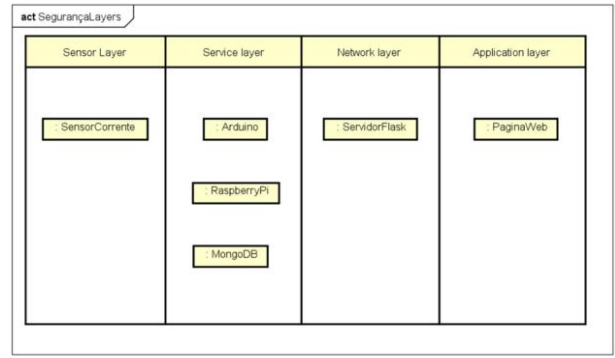

**Fig. 14:** Modelo de segurança do sistema IoT.

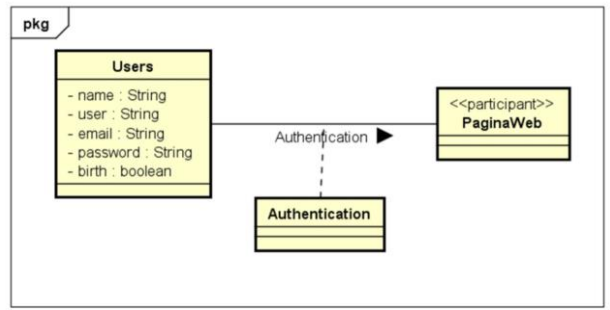

**Fig. 15:** Representação da autenticação do usuário..

b) *Identificação dos objetivos tecnológicos:* Os objetivos tecnológicos representam os requisitos não funcionais, ou seja, relacionados à tecnologia utilizada para realização do objetivo estratégico. Os objetivos tecnológicos são representados com casos de uso de linhas pontilhadas.

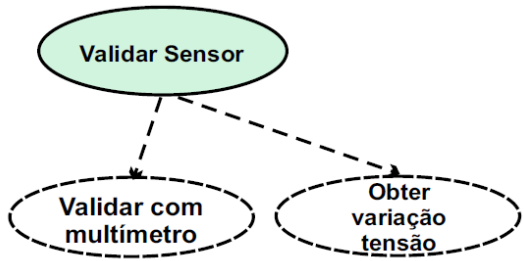

**Fig. 16:** Objetivos tecnológicos para a função "Validar Sensor".

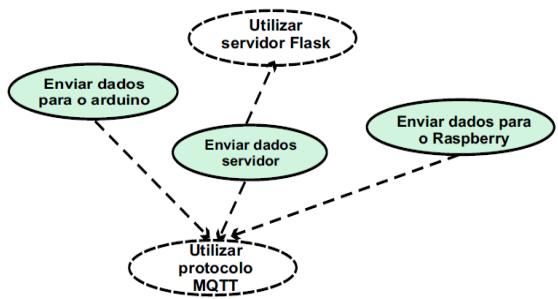

**Fig. 17:** Objetivos tecnológicos para as funções "Enviar dados servidor", "Enviar dados para Raspberry" e "Enviar dados para Arduino".

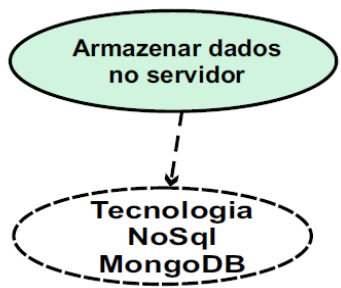

**Fig. 18:** Objetivos tecnológicos para a função "Armazenar dados no servidor".

## Módulo de Previsão do Consumo de Energia Elétrica Utilizando Redes Neurais LSTM para o EnergySaver.

A previsão de demanda de energia elétrica é um componente importante em qualquer sistema de gerenciamento de energia moderno. Uma das técnicas utilizadas atualmente para previsão, são as Redes Neurais Artificiais (RNA). Dentre vários tipos de RNAs, a Long Short-Term Memory (LSTM) é bastante utilizada para a previsão do consumo de energia elétrica, obtendo resultados satisfatórios.

No sistema EnergySaver, esse modelo de previsão com LSTM, foi incrementado para ser executado no computador servidor automaticamente no primeiro dia útil de cada mês. Deste modo, a partir dos dados contidos nos datasets de treinamento e testes, a rede é executada e o resultado é exibido para o usuário no final do processo.

A figura 19 abaixo, apresenta o esquema do módulo, onde os dados recebidos e armazenados no MongoDB, são usados para alimentar a rede LSTM e sua execução automática.

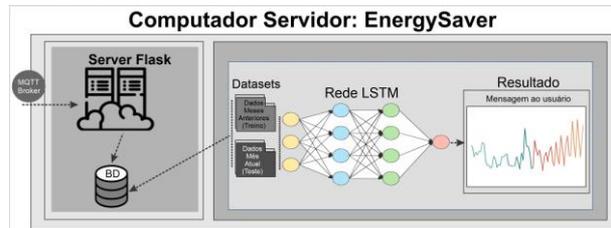

**Fig. 19:** Módulo LSTM.

Na figura 19, os dados enviados pelo RaspberryPi, via protocolo MQTT, são recebidos em tempo real pelo servidor flask e armazenados no MongoDB (DB). Esses dados, formam os Datasets de Treino e Teste, dispostos em formato de planilha eletrônica (.csv).

*Dataset*

Os dados utilizados na pesquisa foram coletados em um prédio de uma universidade pública, considerando o consumo total de energia, nos períodos de janeiro de 2019 a Agosto de 2019.

Após transformar os dados brutos para índice em base fixa, o dataset ficou com 33.830 registros no total. Desse total, foram utilizados para treinamento 29.492 registros referentes aos meses de Janeiro a Julho de 2019, correspondendo a aproximadamente 87% dos registros; e para testes 4.338 registros referentes ao mês de Agosto de 2019, correspondendo a 13% do total dos registros no dataset.

Para melhor compreensão do dataset, foi plotado o gráfico da figura 20 abaixo, onde apresenta a média do consumo diário/mensal no período de Janeiro de 2019 a Agosto de 2019.

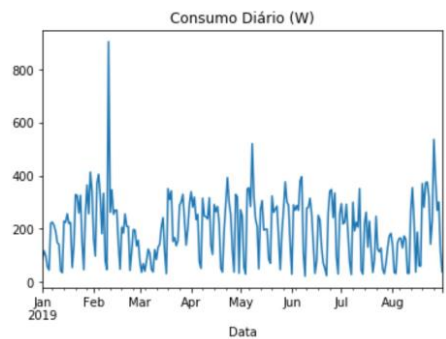

**Fig. 20:** Gráfico do consumo mensal.

Observando a Figura 20, existem algumas variações na carga. Isso provavelmente é relacionado à coleta em dias úteis, com maior consumo e fins de semana onde a demanda é menor. Destaque-se também alguns picos de consumo que fogem um pouco da média.

*Resultados Experimentais utilizando um modelo de rede LSTM*

Inicialmente foram definidas as principais configurações do algoritmo LSTM relacionados a: normalização (0 e 1); os datasets para treinamento e testes; as épocas (100); o batch_size (32); o tamanho da janela de deslizamento (90 tanto para treinamento quanto para teste); e o modelo utilizado foi o sequencial.

Nos experimentos buscou-se prever o consumo de energia elétrica para o mês de Agosto de 2019, utilizando como treinamentos os meses de Janeiro a Julho de 2019. A Figura 21 apresenta os resultados após os testes.

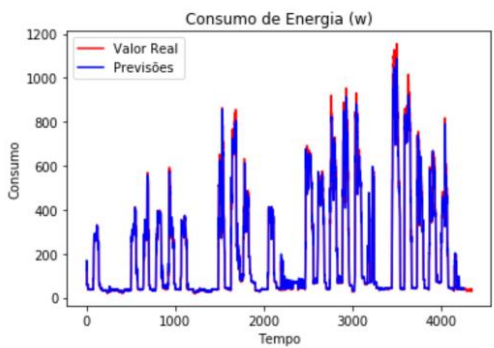

**Fig. 21:** Resultado de previsão nos testes.

A figura 21, mostra que o algoritmo LSTM conseguiu aprender com os dados de treinamento, e quando aplicado nos testes, as previsões (em azul) para o mês de Agosto de 2019 foram próximas do valor real (em vermelho). Pode-se observar que segue a tendência geral da carga futura nos testes.

*Avaliação dos Experimentos e conclusões.*

Também avaliou-se a partir de três critérios padrões comumente usados como métrica de desempenho: root mean squared error (RMSE), mean squared error (MSE) e mean absolute error (MAE), entre valores reais e resultados da previsão, conforme tabela 1 abaixo.

**Tabela 1:** Resultados médios MSE, RMSE e MAE nos Testes.

| Algoritmo | MSE | RMSE | MAE |
|---|---|---|---|
| LSTM | 0.0017 | 0.041 | 0.0263 |

Na Tabela 1, a medidas MSE, RMSE e MAE que se baseiam no erro percentual de previsão. A MSE é uma função calcula a média dos erros do modelo ao quadrado, ou seja, diferenças menores têm menos importância, enquanto diferenças maiores recebem mais peso. No mesmo sentido, a função RMSE é calculada a partir da Raiz quadrada da MSE. Já a MAE toma o valor absoluto entre a previsão do modelo, e o valor real, ao quadrado. Atribui o mesmo peso a todas as diferenças, de maneira linear.

De modo geral, quanto mais próximo de zero (0), melhor o resultado da rede, ou seja, os valores obtidos pelo algoritmo nos testes, foi bem próximo dos valores reais de consumo existente no dataset.

Assim sendo, pode-se concluir que o módulo LSTM para previsão de consumo de energia elétrica, ao ser integrado ao sistema energy Saver, possibilitará um monitoramento completo de todo o processo, desde a coleta do consumo do equipamento individual, até o processo de previsão para o mês seguinte, de forma automatizada.

## Considerações sobre o sistema e equipe

*Considerações sobre o sistema*

A versão 1.1 do sistema Energy Saver para monitoramento do consumo de energia elétrica trouxe uma abordagem muito importante para os dias atuais, tendo em vista tratar diretamente da eficiência energética.

Isso porque a partir desses dados obtidos, pode-se direcionar ações que possibilitem a redução do consumo de energia elétrica.

De acordo com US National Policy Development Group, a eficiência energética é a capacidade de utilizar menos energia para produzir a mesma quantidade de iluminação, aquecimento, transporte e outros serviços baseados na energia.

Deste modo, melhorar o desempenho energético pode trazer benefícios importantes para as organizações, tanto na questão ambiental quanto às providências ligadas à redução do consumo energético e aumento da sua eficiência, promovendo assim o desenvolvimento sustentável.

Nesse sentido, os passos futuros deste projeto serão:
a) Medição do consumo de eletricidade em sistemas de ar-condicionado do LABIC;
b) Aplicações de Técnicas de Deep Learning na análise do consumo de energia;

Por fim, a próxima versão do sistema buscará tanto o monitoramento do consumo energético de sistemas de ar-condicionado quanto algumas ações a partir de técnicas de Deep Learning para promover a redução desse consumo.